\documentclass{article}

\usepackage[utf8]{inputenc}
\usepackage{geometry}
\usepackage{amsmath, amssymb, amsfonts}
\usepackage{graphicx}
\usepackage{hyperref}
\usepackage{url}
\usepackage{algorithm}
\usepackage{algorithmic}
\usepackage{booktabs}
\usepackage{multirow}
\usepackage{caption}
\usepackage{listings}
\usepackage{xcolor}

\lstdefinestyle{dalia}{
  basicstyle=\ttfamily\footnotesize,
  frame=single,
  rulecolor=\color{black!30},
  backgroundcolor=\color{black!5},
  breaklines=true,
  tabsize=2,
  showstringspaces=false,
  captionpos=b
}

\hypersetup{
    colorlinks=true,
    linkcolor=blue,
    citecolor=blue,
    urlcolor=blue
}

\title{
\textbf{Towards a Declarative Agentic Layer for Intelligent Agents in MCP-Based Server Ecosystems}
}

\author{
  María Jesús Rodríguez-Sánchez \\
  Universidad de Granada \\
  \texttt{mjesusrodriguez@ugr.es}
  \and
  Manuel Noguera \\
  Universidad de Granada \\
  \texttt{mnoguera@ugr.es}
  \and
  Angel Ruiz-Zafra \\
  Universidad de Granada \\
  \texttt{angelr@ugr.es}
  \and
  Kawtar Benghazi \\
  Universidad de Granada \\
  \texttt{benghazi@ugr.es}
}

\date{} 

\begin{document}

\maketitle

\begin{abstract}
Recent advances in Large Language Models (LLMs) have enabled the development of increasingly complex agentic and multi-agent systems capable of planning, tool use and task decomposition. However, empirical evidence shows that many of these systems suffer from fundamental reliability issues, including hallucinated actions, unexecutable plans and brittle coordination. Crucially, these failures do not stem from limitations of the underlying models themselves, but from the absence of explicit architectural structure linking goals, capabilities and execution.

This paper presents a declarative, model-independent architectural layer for grounded agentic workflows that addresses this gap. The proposed layer, referred to as DALIA (Declarative Agentic Layer for Intelligent Agents), formalises executable capabilities, exposes tasks through a declarative discovery protocol, maintains a federated directory of agents and their execution resources, and constructs deterministic task graphs grounded exclusively in declared operations. By enforcing a clear separation between discovery, planning and execution, the architecture constrains agent behaviour to a verifiable operational space, reducing reliance on speculative reasoning and free-form coordination.

We present the architecture and design principles of the proposed layer and illustrate its operation through a representative task-oriented scenario, demonstrating how declarative grounding enables reproducible and verifiable agentic workflows across heterogeneous environments.
\end{abstract}


\section{Introduction}


Recent advances in Large Language Models (LLMs) have reshaped the landscape of artificial intelligence. Contemporary systems, including widely adopted conversational agents and open foundational models, have demonstrated strong performance across a wide range of cognitively demanding tasks. Peer-reviewed studies show that LLMs can integrate and synthesize information from multiple sources \cite{zhang2024multidoc} and interact with external tools through structured interfaces \cite{schick2024toolformer}. These developments have expanded the role of LLMs beyond text generation and towards general-purpose cognitive engines capable of supporting complex reasoning, automating multi-step workflows and engaging in broader computational tasks \cite{li-etal-2024-fundamental}.

The growing need to make LLMs more controllable, structured and goal-directed has led to the emergence of \textit{agentic AI}, a paradigm that integrates planning, memory, tool use and interaction into LLM-driven systems \cite{schneider2024agentic, ieee-agentic-survey}. Recent surveys emphasize that agentic architectures arise because conventional LLMs, despite their impressive capabilities, rely heavily on linguistic inference and operate without explicit grounding in the actions they are able to perform \cite{he2025multiagent}. As a consequence, LLM-based agents often produce unvalidated or infeasible behaviours, including attempting operations they cannot execute or deviating from task objectives. This limitation becomes particularly evident in the construction of \textit{task graphs}, a common abstraction for decomposing complex problems into executable steps \cite{llm-taskgraphs}. Prior work shows that these graphs are frequently either hard-coded or generated without awareness of the real capabilities available in the environment, resulting in incoherent routes, hallucinated operations and unexecutable workflows \cite{he2025multiagent}.

In parallel, the Model Context Protocol (MCP) has emerged as a lightweight standard for enabling LLMs to discover and invoke external tools. MCP provides a unified and practical interface for tool execution \cite{mcp-spec}, but it remains semantically limited: it exposes a flat list of tools without relationships, capability semantics, task-level structure or support for multi-server coordination. Recent analyzes of MCP and its extensions recognize the need to enrich the protocol in order to support more expressive and scalable agentic systems \cite{scalemcp}. As a consequence, agents lack the structured information required to reason about which operations are feasible, how tools relate to each other or how to compose them into executable tasks.

These observations reveal a critical missing component in current agentic ecosystems: a declarative layer that connects goals, capabilities and execution in a structured and operationally grounded manner. To address this gap, we introduce \textbf{DALIA}, a \textit{Declarative Agentic Layer for Intelligent Agents} that provides: (1) a formal semantic model of capabilities that enriches tool definitions with functional categories, roles, preconditions, postconditions and compositional constraints; (2) an Agentic Task Discovery Protocol (ATDP) that expresses goals and subtasks declaratively, enabling agents to determine what they can accomplish based on real capabilities rather than inferred assumptions; (3) a federated directory for MCP servers that supports capability discovery, metadata sharing and multi-server interoperability; and (4) a deterministic, capability-grounded planner that synthesizes executable and verifiable task graphs aligned with the actual operations available in the environment. Together, these components provide the structural foundation needed to build reliable, reproducible and verifiable agentic systems.

The remainder of this paper is organised as follows: Section \ref{background} reviews the background and motivation underlying current limitations in agentic and multi-agent systems, with particular attention to task graphs, capability grounding and MCP-based infrastructures. Section \ref{proposal} introduces the proposed declarative agentic layer, detailing its architectural principles and core components, including the capability semantic model, the Agentic Task Discovery Protocol and the federated Agent Directory. Section \ref{example} illustrates the operation of the proposed architecture through a representative task-oriented scenario, demonstrating how grounded and reproducible agentic workflows are constructed and executed. Finally, Section \ref{conclusion} discusses the implications of this approach, outlines current limitations and highlights directions for future research.
\section{Background and Motivation}
\label{background}

LLM-based multi-agent systems (MAS) have expanded rapidly across domains such as software engineering, scientific problem solving and general task automation, with frameworks like MetaGPT \cite{hong2023metagpt}, ChatDev \cite{qian2023chatdev} and AgentVerse \cite{chen2023agentverse} illustrating the growing interest in distributing complex tasks among specialised agents. Yet empirical evidence consistently questions their reliability. The most comprehensive evaluation to date, by Cemri et al., examines 1,642 executions across seven MAS frameworks and reports failure rates between 41\% and 86\% \cite{cemri2025mast}. Through a Grounded Theory analysis, the authors introduce MAST, the first empirical taxonomy of MAS failure modes, identifying fourteen recurrent errors across three categories: system-design flaws (FC1), agent misalignment (FC2) and deficient verification (FC3). Crucially, these failures stem not from deficiencies in the underlying models but from a persistent lack of structure in MAS organization, communication and validation. Complementary studies reinforce this diagnosis: manually engineered MAS tend to be brittle and domain-bound, while LLM-generated systems such as AFlow \cite{zhang2024aflow} or MAS-GPT \cite{ye2025masgpt} often produce incoherent or unexecutable agent structures. Recent surveys further highlight that current MAS rely heavily on linguistic improvisation rather than explicit representations of roles, capabilities, or workflows \cite{schneider2024agentic, sapkota2025taxonomy}. Taken together, these findings show that MAS require structural redesigns and explicit architectural support rather than additional prompting or fine-tuning, motivating the search for principled mechanisms that provide coordination, grounding and reliability.

A second line of work investigates the use of task graphs as a way to impose structure on multi-agent workflows. CODER \cite{chen2024coder} exemplifies this direction by coordinating specialized agents through a predefined JSON-based task graph, reducing common problems such as loops, inconsistent decisions, and loss of context. However, prior work demonstrates that LLMs cannot reliably generate task graphs on their own: graphs produced through unconstrained language modelling tend to be incomplete, incoherent, or ungrounded in the capabilities available in the environment \cite{llm-taskgraphs, he2025multiagent}. Although hard-coded graphs provide stability, they lack adaptability and cannot support dynamic task discovery. These limitations indicate that task graphs alone are insufficient unless anchored to a declarative representation of real capabilities.

In parallel, the Model Context Protocol (MCP) has emerged as a lightweight standard for exposing tools to LLMs. MCP servers provide a minimal interface consisting of tools, resources, and prompts, along with basic discovery operations such as \texttt{list\_tools} and \texttt{list\_resources}. Despite its usefulness as a transport layer, the protocol remains structurally under-specified: tools lack semantic descriptions, dependencies, roles or relations; resources are exposed without indications of relevance or compatibility; and servers are fully isolated, with no mechanism for federation or multi-server coordination. MCP offers no representation of tasks, workflows or execution constraints, placing the entire burden of planning on the LLM. Extensions such as ScaleMCP \cite{hou2025mcp} and MCPEval \cite{liu2025mcpeval} improve operational dimensions such as synchronisation and evaluation but do not address the absence of semantics, composition or task-level structure.

\begin{table}[t]
\centering
\footnotesize
\begin{tabular}{p{3.3cm} p{4.2cm} p{4.2cm}}
\toprule
\textbf{Approach} & \textbf{Identified Limitations} & \textbf{Uncovered Needs} \\
\midrule
MAST (Cemri et al.) \cite{cemri2025mast} & High failure rates; structural flaws in design, alignment and verification. & Explicit architectural support for coordination, grounding and validation. \\
\hline
CODER \cite{chen2024coder} & Predefined task graphs; LLM-generated graphs incoherent or ungrounded \cite{llm-taskgraphs}. & Grounded, adaptable task planning tied to real capabilities. \\
\hline
AFlow \cite{zhang2024aflow} / MAS-GPT \cite{ye2025masgpt} & Manually designed MAS brittle; LLM-generated MAS incoherent or unexecutable. & Formal representation of roles, capabilities and workflow constraints. \\
\hline
MCP standard \cite{hasan2025mcpfirstglance} & Tools lack semantics, relations or dependencies; no tasks; no federation. & Declarative capability model; workflow structure; multi-server metadata. \\
\hline
ScaleMCP \cite{hou2025mcp} / MCPEval \cite{liu2025mcpeval} & Operational improvements only; no semantics, tasks or composition. & Higher-level layer for task discovery and capability reasoning. \\
\hline
Agentic AI (current approaches) \cite{schneider2024agentic, sapkota2025taxonomy, acharya2025agentic} & Capabilities inferred; tasks unvalidated; coordination via free-form dialogue. & Principled declarative layer linking goals, capabilities and execution.\\
\bottomrule
\end{tabular}
\caption{Comparison of limitations in prior work and unmet needs addressed by DALIA.}
\label{tab:comparison}
\end{table}

These challenges align with the broader diagnosis presented in recent Agentic AI surveys, which emphasise that although the field aspires to build systems capable of reasoning, planning, remembering, acting and interacting, current approaches rely almost entirely on LLM-driven inference rather than grounded operational models \cite{schneider2024agentic, acharya2025agentic}. Capabilities are inferred rather than declared, tasks are generated but not validated, tools are used but not described and coordination emerges from free-form dialogue rather than from formal protocols. The combined limitations of MAS frameworks, task-graph approaches, MCP infrastructure and Agentic AI methodologies are summarised in Table~\ref{tab:comparison}. Taken together, these findings expose a structural gap in the current agentic AI ecosystem. Existing approaches lack a declarative layer that explicitly links goals, capabilities and execution, leaving agents to operate through inference rather than grounded operational knowledge. Addressing this limitation requires an architecture able to (1) describe capabilities formally, (2) discover and organise tasks in a structured manner, (3) federate heterogeneous servers and (4) generate deterministic task plans aligned with real executable operations. These requirements motivate the development of \textbf{DALIA}, a Declarative Agentic Layer for Intelligent Agents, introduced in the next section as a foundation for reliable, reproducible and verifiable agentic systems.

\section{DALIA Architecture}
\label{proposal}

DALIA is a declarative architectural layer designed to support the construction of reliable and controllable agentic systems. Rather than introducing a new agent model or interaction paradigm, DALIA provides an intermediate layer that explicitly links goals, capabilities and execution. This layer operates independently of the underlying Large Language Models and complements existing agentic frameworks by supplying the structural information required for grounded planning, coordination and verification.

\begin{figure}[t]
  \centering
  \includegraphics[width=\linewidth]{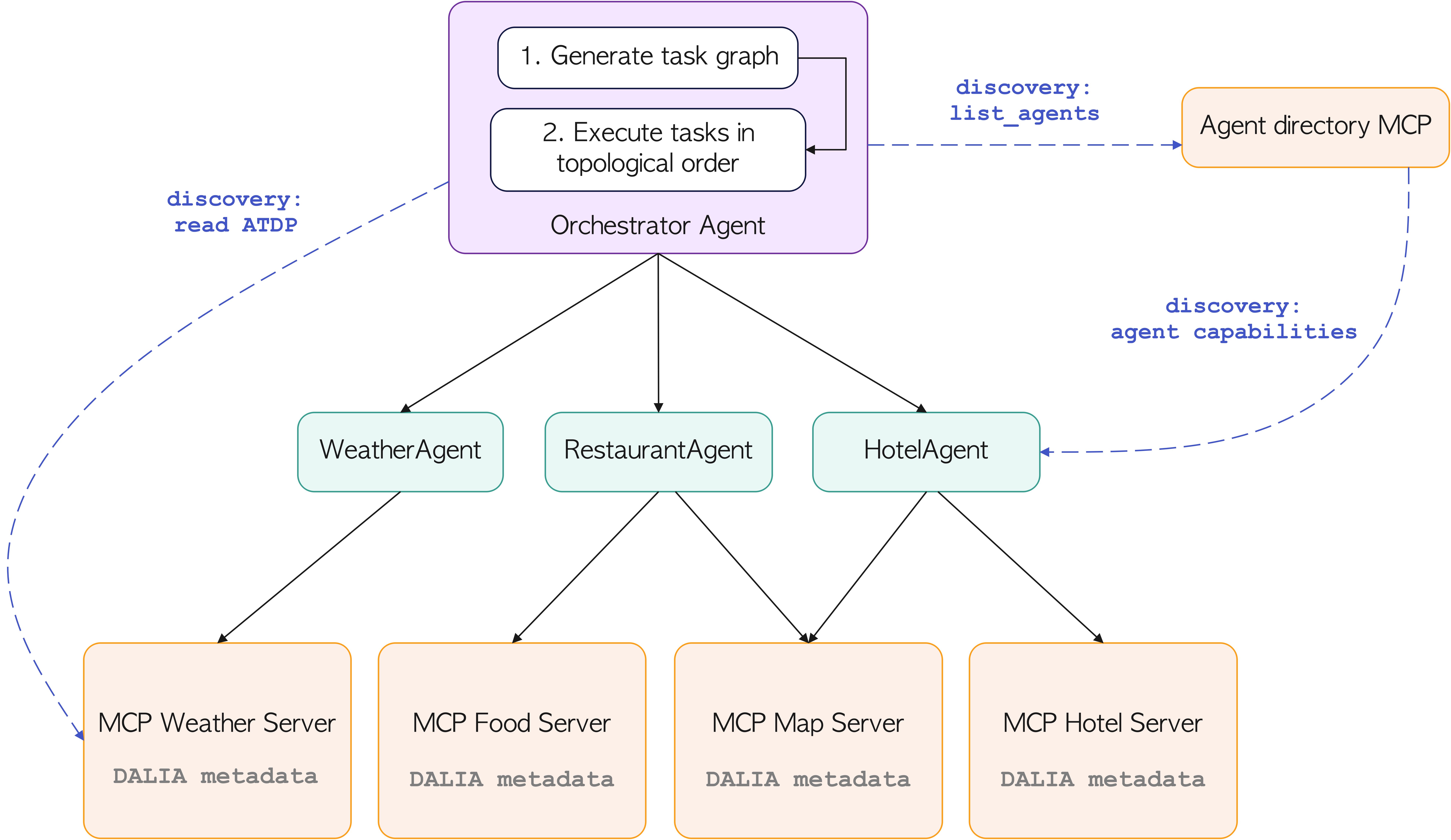}
  \caption{Overview of the DALIA architecture.}
  \label{fig:dalia-architecture}
\end{figure}

At a high level, DALIA sits between LLM-driven agents and a heterogeneous ecosystem of tools, services and execution environments, as illustrated in Figure~\ref{fig:dalia-architecture}. Agents remain responsible for reasoning, interpretation and decision-making, while DALIA exposes a shared declarative substrate that defines what actions are available, how they can be composed and under which conditions they can be executed. This separation of concerns enables agents to operate over an explicit and structured space of capabilities, instead of relying on implicit assumptions inferred from natural language alone.

The architecture of DALIA is organised around a small set of explicit additions to existing agentic infrastructures: (1) a declarative capability representation that enriches MCP-exposed tools and services with structured semantic descriptors; (2) the Agentic Task Discovery Protocol (ATDP), which provides a declarative interface for exposing available tasks, inputs and outputs; (3) a federated Agent Directory MCP that maintains a global view of available agents and the executable capabilities exposed by the MCP servers they can access; and (4) a deterministic task orchestration mechanism that constructs and executes task graphs grounded exclusively in declared capabilities.

\begin{lstlisting}[style=dalia, caption={Example agent description stored in the Agent Directory MCP}, label={lst:agent-directory-example}]
{
  "agent_id": "RestaurantAgent",
  "role": "booking_agent",
  "mcp_servers": ["restaurant_mcp"],
  "executable_capabilities": [
    "restaurant.search",
    "restaurant.reserve"
  ]
}
\end{lstlisting}

Building on this representation, DALIA introduces the \textit{Agentic Task Discovery Protocol (ATDP)}, which provides a declarative interface for exposing the tasks supported by each MCP server, together with their required inputs, outputs and domain annotations. The orchestrator uses ATDP to determine which tasks can be performed in a given environment and to select suitable agents and servers without generating speculative or ungrounded plans. Capability discovery is further supported by a federated \textit{Agent Directory MCP}, which maintains a global view of available agents, their declared capabilities and the MCP servers they can access.

To make this association explicit, the Agent Directory MCP records, for each agent, the set of MCP servers it can access and the corresponding executable capabilities derived from those servers. The directory does not replicate capability definitions; instead, it maintains a structured index that allows the orchestration layer to determine which agents can execute which capabilities across the federated environment. Listing~\ref{lst:agent-directory-example} illustrates an example of an agent entry in the Agent Directory MCP.

DALIA further operationalises these elements through a deterministic task orchestration mechanism. Rather than relying on free-form agent coordination, the orchestrator constructs task graphs exclusively from declared capabilities and executes them in topological order, ensuring that each step corresponds to a valid and executable operation. This design enforces a strict separation between discovery and execution, providing reproducibility, verifiability and robustness across agentic workflows.

By externalising capability descriptions, task discovery and planning from the agents themselves, DALIA addresses several of the structural limitations identified in existing agentic systems. Agents are no longer required to infer the existence or applicability of actions through linguistic reasoning alone; instead, they operate within a declarative space that constrains and guides their behaviour. This design reduces the likelihood of hallucinated actions, invalid task graphs and unexecutable workflows, while enabling systematic verification and reuse of agentic behaviours.

The remainder of this section describes each component of DALIA in more detail. We first introduce the capability semantic model, followed by the Agentic Task Discovery Protocol and the federated MCP directory, and then present the task planning mechanism that operationalises these elements into executable workflows.

\subsection{Capability Semantic Model}

At the core of DALIA lies a declarative semantic model for representing executable capabilities exposed by agents, tools and services. Rather than treating operations as opaque functions described solely through textual prompts or function signatures, DALIA models capabilities as first-class architectural entities with explicit, machine-interpretable structure. This semantic model establishes the foundation for grounded task discovery, planning and execution across heterogeneous agentic environments.

In DALIA, a capability represents a concrete operational unit that can be invoked through an MCP server. Each capability corresponds to a real, executable action supported by the system and is described using a structured set of semantic attributes that specify not only how the operation can be called, but also what it represents and under which conditions it is applicable. At a minimum, these attributes capture the functional role of the capability, its domain or category, the required inputs and the produced outputs. The model can optionally include additional constraints such as preconditions, postconditions or compatibility relations with other capabilities. By externalising this information from natural language prompts into explicit declarations, DALIA enables systematic reasoning about feasibility, compatibility and composition prior to execution.

\begin{lstlisting}[style=dalia, caption={Illustrative example of a capability declaration following the DALIA semantic model}, label={lst:capability-example}]
{
  "capability_id": "restaurant.search",
  "role": "information_retrieval",
  "domain": "food",
  "inputs": ["location", "date", "party_size"],
  "outputs": ["restaurant_list"],
  "preconditions": ["location_known"],
  "postconditions": ["results_available"]
}
\end{lstlisting}

The capability semantic model is designed to remain independent of any specific agent implementation or underlying LLM. In line with prior work on tool-augmented language models, capabilities in DALIA are declared externally to the model and exposed through structured discovery interfaces, allowing agents and orchestrators to reason over available actions without relying on model-specific assumptions \cite{schick2024toolformer}. Within DALIA, these capability declarations are provided locally by MCP servers as part of their configuration, following the DALIA semantic model, and exposed through MCP-compatible discovery mechanisms. This enables uniform consumption of capability descriptions by different agents and orchestrators, without requiring direct knowledge of server implementations or internal tool logic. Figure~\ref{lst:capability-example} illustrates an example of a capability declaration following this semantic model.

The capability semantic model specifies the structure and semantics of capability descriptions without mandating a fixed ontology or domain-specific vocabulary. Instead, it defines a minimal and extensible schema that accommodates diverse application domains while preserving interoperability across servers. This design allows DALIA to operate in dynamic environments where capabilities may evolve, be added or removed over time, without requiring changes to agent logic or orchestration strategies.

The following subsection builds on this semantic foundation by introducing the Agentic Task Discovery Protocol, which leverages declared capabilities to expose tasks and goals in a structured and discoverable manner.

\subsection{ATDP: Agentic Task Discovery Protocol}

While the capability semantic model defines what operations are executable in the system, DALIA requires an explicit mechanism to express what can be achieved by combining those operations. To this end, DALIA introduces the \textit{Agentic Task Discovery Protocol (ATDP)}, a declarative protocol that exposes tasks and goals supported by an environment in terms of the capabilities required to accomplish them.

In DALIA, a task represents a higher-level objective that can be fulfilled through the execution of one or more declared capabilities. Tasks are not generated speculatively by agents, nor inferred implicitly from natural language descriptions. Instead, they are declared explicitly by MCP servers using ATDP, together with structured metadata that specifies their intent, required inputs, expected outcomes and associated capabilities. This allows agents and orchestrators to determine which tasks are feasible in a given environment before attempting to plan or execute them.

ATDP operates as a discovery layer built on top of the capability semantic model. MCP servers expose the set of tasks they support through ATDP-compatible discovery mechanisms, enabling agents to query available goals in a structured and machine-interpretable manner. Each task declaration references the capabilities it depends on, making explicit the relationship between high-level objectives and low-level executable operations. As a result, task discovery in DALIA is grounded in real system capabilities rather than in assumptions derived from linguistic reasoning.

Listing~\ref{lst:task-example} illustrates an example of a task declaration following ATDP. The example shows how a task is described in terms of its identifier, intent, required inputs, produced outputs and the set of capabilities that may be composed to achieve it. This representation enables the orchestrator to reason about task feasibility, select appropriate agents and construct valid task plans without relying on ad hoc heuristics or prompt-based decomposition.

\begin{lstlisting}[style=dalia, caption={Illustrative example of a task declaration using the Agentic Task Discovery Protocol (ATDP)}, label={lst:task-example}]
{
  "task_id": "restaurant.booking",
  "intent": "book_restaurant",
  "inputs": ["location", "date", "party_size"],
  "outputs": ["booking_confirmation"],
  "capabilities": [
    "restaurant.search",
    "restaurant.reserve"
  ]
}
\end{lstlisting}

Importantly, ATDP does not prescribe a fixed task taxonomy or a predefined decomposition strategy. Instead, it provides a minimal declarative structure for exposing tasks while preserving flexibility across domains and application scenarios. By separating task discovery from task planning and execution, ATDP allows DALIA to support dynamic environments in which available goals may evolve alongside underlying capabilities.

The next subsection introduces the deterministic task orchestration mechanism, which leverages declared capabilities and ATDP-exposed tasks to construct and execute grounded task graphs.

\subsection{Agent Directory MCP}

While the capability semantic model and ATDP define what operations are executable and which tasks are achievable, DALIA requires an explicit mechanism to describe \emph{who} can execute those operations. To this end, DALIA introduces a federated \textit{Agent Directory MCP}, a declarative registry that maintains a global view of available agents, their roles and the MCP servers they can access.

In DALIA, agents are treated as execution entities rather than autonomous planners. Each agent is described declaratively in the Agent Directory MCP using structured metadata that specifies its identifier, functional role, supported domains and accessible MCP servers. Importantly, the directory does not duplicate capability definitions. Instead, it links agents to the MCP servers that expose capabilities following the DALIA semantic model, making explicit which agents are authorised to execute which sets of operations.

This design decouples agent identity and responsibility from capability specification. Agents do not embed knowledge about available tools or tasks internally; rather, they act as clients of MCP servers and rely on the Agent Directory to expose which execution resources they can use. The orchestrator queries the Agent Directory MCP during the discovery phase to determine which agents are suitable for executing the capabilities required by a given task graph.

Listing~\ref{lst:agent-declaration-example} illustrates an example of an agent declaration maintained by the Agent Directory MCP. The example shows how an agent is described in terms of its role, supported domains and the MCP servers it can access, without embedding task logic or capability semantics within the agent itself.

\begin{lstlisting}[style=dalia, caption={Illustrative example of an agent declaration stored in the Agent Directory MCP}, label={lst:agent-declaration-example}]
{
  "agent_id": "RestaurantAgent",
  "role": "task_executor",
  "domains": ["food"],
  "accessible_servers": [
    "mcp_food_server",
    "mcp_map_server"
  ]
}
\end{lstlisting}

The Agent Directory MCP is designed to be federated and extensible. Multiple directories may coexist across environments, and agents can be added, removed or updated dynamically without requiring changes to capability definitions, task specifications or orchestration logic. By externalising agent availability and execution permissions into a declarative layer, DALIA enables systematic reasoning about agent selection, execution responsibility and multi-agent coordination.

Together with the capability semantic model and ATDP, the Agent Directory MCP completes the declarative foundation of DALIA, allowing task discovery, agent selection and execution planning to be grounded in explicit and verifiable system descriptions rather than inferred assumptions.

The next subsection describes how these declarative elements are operationalised through deterministic task orchestration.

\subsection{Deterministic Task Orchestration}

While the capability semantic model and ATDP define which operations are executable and which tasks are achievable in a given environment, DALIA requires a concrete mechanism to transform this declarative knowledge into actual executions. To this end, DALIA introduces a deterministic task orchestration component responsible for constructing and executing task graphs grounded exclusively in declared capabilities and explicitly discovered tasks.

Task orchestration in DALIA is performed by an external orchestration layer that operates independently from individual agents. This orchestrator consumes three sources of declarative information: (i) capability descriptions exposed by MCP servers following the DALIA semantic model, (ii) task declarations exposed through ATDP, and (iii) agent metadata obtained from the federated Agent Directory MCP. Based on this information, the orchestrator synthesises executable task graphs in which each node corresponds to the execution of a specific declared capability by a designated agent, and each edge represents an explicit execution dependency derived from declared inputs, outputs and constraints.

Crucially, task graphs in DALIA are not generated through unconstrained language modelling, nor inferred implicitly from prompts or conversational reasoning. Instead, graph construction is deterministic with respect to the available declarative information: only capabilities that are explicitly declared, discoverable and compatible are considered during planning. As a result, the orchestrator never constructs plans that reference unavailable actions, incompatible operations or unsupported goals. This stands in contrast to many existing agentic systems, where planning and execution are interleaved and frequently rely on speculative reasoning or on-the-fly correction.

The separation between discovery, planning and execution is a core design principle of DALIA. Capability and task discovery are completed prior to orchestration, ensuring that planning operates over a closed and verifiable set of executable operations. Once a task graph has been constructed, execution proceeds in a controlled and reproducible manner: capabilities are invoked in topological order, intermediate results are propagated explicitly between nodes according to declared inputs and outputs, and failures can be detected and handled deterministically. Because each execution step corresponds to a concrete operation with well-defined preconditions and outcomes, task graphs can be systematically verified, debugged and reused across executions and environments.

\begin{figure}[t]
  \centering
  \includegraphics[width=\linewidth]{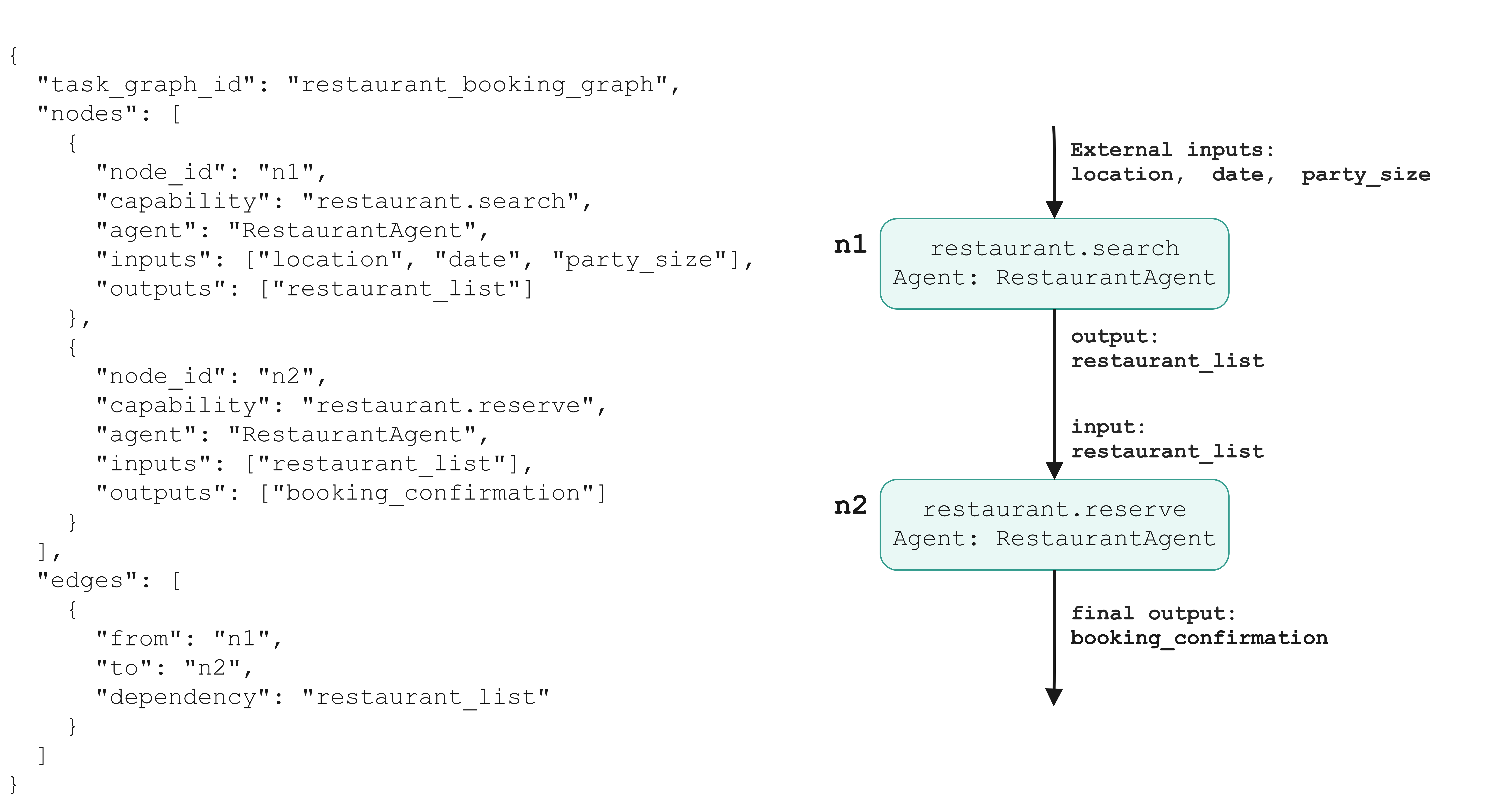}
  \caption{Example of a deterministic task graph constructed from declared capabilities and ATDP-exposed tasks.}
  \label{fig:task-graph-example}
\end{figure}

Figure~\ref{fig:task-graph-example} illustrates an example of a task graph constructed by DALIA for a simple restaurant booking scenario. The example highlights how high-level tasks exposed through ATDP are decomposed into a sequence of capability executions, each assigned to a specific agent and connected through explicit data dependencies. This representation makes the execution logic explicit, inspectable and independent of any particular LLM or agent implementation.

By externalising orchestration logic from individual agents and grounding execution in declarative capability and task descriptions, DALIA reduces reliance on free-form agent coordination and linguistic improvisation. Agents act primarily as execution endpoints for declared capabilities, while DALIA provides the structural guarantees required for reliable, reproducible and verifiable agentic workflows.

\subsection{DALIA Execution Pipeline}

While the previous subsections describe the architectural components of DALIA in isolation, their interaction during execution follows a well-defined and structured pipeline. DALIA organises agentic execution as a sequence of explicit phases that separate discovery, planning and execution, ensuring that each step operates over a grounded and verifiable representation of system capabilities.

At a high level, a DALIA-enabled interaction proceeds through the following phases. First, a high-level goal or user request is received by the orchestration layer, without assuming any specific decomposition strategy. Second, the orchestrator performs capability and task discovery by querying MCP servers and ATDP endpoints, obtaining a closed set of executable operations and supported tasks. Third, the Agent Directory MCP is consulted to determine which agents are eligible to execute the required capabilities, based on their declared access to MCP servers and execution permissions.

Once discovery is complete, the orchestrator synthesises a task graph that satisfies the requested goal using only declared capabilities, tasks and agent assignments. This planning phase operates over explicit structural information rather than free-form language inference. Finally, the constructed task graph is executed in a controlled manner, invoking capabilities in topological order and propagating intermediate results between execution steps according to declared inputs and outputs.

Importantly, DALIA enforces a strict separation between these phases. Discovery is completed prior to planning, and planning is completed prior to execution. This design prevents agents from attempting unsupported actions, generating speculative plans or invoking unavailable tools at runtime. The internal reasoning mechanisms used to guide task selection or graph construction are intentionally left unspecified, allowing DALIA to remain compatible with rule-based, heuristic or LLM-driven orchestration strategies.

This phased execution pipeline provides a clear operational semantics for DALIA while preserving flexibility in implementation. The following section illustrates this pipeline through a concrete end-to-end scenario.
\section{Illustrative Scenario}
\label{example}

\begin{figure}[t]
  \centering
  \includegraphics[width=\linewidth]{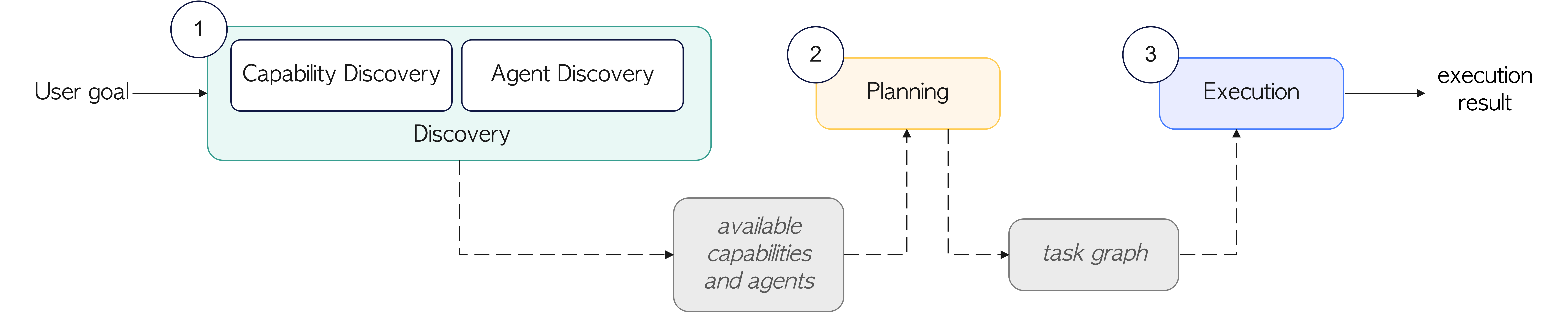}
  \caption{DALIA execution pipeline organised into discovery, planning and execution phases.}
  \label{fig:dalia-pipeline}
\end{figure}

To illustrate how DALIA operates in practice, we consider a simplified restaurant booking scenario that follows the execution pipeline depicted in Figure~\ref{fig:dalia-pipeline}. The purpose of this example is not to demonstrate optimisation or performance, but to clarify how goals, capabilities, agents and execution are linked through DALIA’s declarative architecture.

\paragraph{Phase 1: Discovery}
The process begins when a user expresses a high-level goal, such as \emph{“book a restaurant for four people tomorrow in the city centre”}. This request is treated as an abstract objective and is not immediately decomposed into actions or execution steps.

During the discovery phase, DALIA constructs an explicit execution context by identifying what can be executed and who can execute it. This phase comprises two complementary sub-steps.
\begin{enumerate}
    \item Capability discovery is performed by querying the available MCP servers and their ATDP endpoints. In this scenario, MCP servers expose declarative descriptions of capabilities such as \texttt{restaurant.search} and \texttt{restaurant.reserve}, together with an ATDP task declaration for \texttt{restaurant.booking} specifying required inputs, expected outputs and associated capabilities.
    
    \item Agent discovery is performed by querying the federated Agent Directory MCP. The directory indicates which agents are authorised to access the MCP servers exposing the discovered capabilities. In this case, a \texttt{RestaurantAgent} is identified as having access to the MCP server providing restaurant-related capabilities.
\end{enumerate}

The result of the discovery phase is a closed and verifiable view of the available executable capabilities and eligible agents, independent of any agent-specific reasoning or assumptions. No agent-level reasoning or prompt-based interpretation is involved at this stage; discovery relies exclusively on the declarative metadata exposed by MCP servers and the Agent Directory.

\paragraph{Phase 2: Planning}
Based on the execution context produced during discovery, DALIA proceeds to the planning phase. The orchestration layer synthesises a task graph that fulfils the user goal using only declared capabilities and authorised agents.

In this scenario, the resulting task graph consists of two nodes corresponding to the execution of \texttt{restaurant.search} followed by \texttt{restaurant.reserve}, connected by an explicit data dependency on the restaurant list produced by the first operation. Each node is assigned to the \texttt{RestaurantAgent}, as determined during agent discovery.

Task graph construction is deterministic with respect to the declarative information available at this stage. Only executable, compatible and explicitly declared operations are considered, ensuring that the resulting plan is valid and inspectable prior to execution. At this point, the task graph fully specifies the execution order, data dependencies and agent assignments, requiring no further interpretation during execution.

\paragraph{Phase 3: Execution}
In the final phase, the task graph is executed in a controlled and reproducible manner. Capabilities are invoked in topological order, and intermediate results are propagated explicitly between nodes according to declared inputs and outputs.

Because each execution step corresponds to a concrete operation with well-defined semantics, failures can be detected and handled deterministically. No additional discovery or speculative planning occurs during execution; the orchestrator strictly follows the previously constructed task graph.

By structuring the interaction in this way, DALIA ensures that goals, capabilities and execution responsibilities remain explicitly separated and traceable throughout the process. All decisions are grounded in declared metadata rather than inferred from natural language, resulting in workflows that are inspectable, reproducible and independent of any specific LLM or agent implementation.
\section{Discussion and Future Directions}
\label{conclusion}

This work introduces DALIA as a declarative architectural layer aimed at addressing structural limitations observed in current agentic and multi-agent systems. Rather than proposing new learning mechanisms, prompting strategies or agent behaviours, DALIA focuses on an often overlooked aspect of agentic AI: the absence of explicit, machine-interpretable structure connecting goals, capabilities, agents and execution. By externalising these elements into a declarative layer, DALIA reframes agentic reasoning as an operation over verifiable system descriptions rather than implicit assumptions derived from natural language.

A central design choice in DALIA is the strict separation between discovery, planning and execution. This separation contrasts with many existing agentic systems, where these phases are tightly interwoven and driven by free-form linguistic reasoning. While such approaches offer flexibility, they also introduce brittleness, unexecutable plans and coordination failures. DALIA demonstrates that introducing explicit declarative grounding can significantly constrain agent behaviour without limiting expressiveness, enabling workflows that are inspectable, reproducible and independent of specific LLM implementations.

It is important to note that DALIA does not prescribe how task graphs are generated internally, nor does it mandate a particular planning algorithm or reasoning mechanism. In practice, different implementations may rely on symbolic planners, heuristic search, rule-based systems or LLM-assisted reasoning to construct task graphs from declarative descriptions. The key contribution of DALIA lies in defining the architectural conditions under which such planning occurs, ensuring that any generated plan remains grounded in declared capabilities, authorised agents and executable operations.

Several directions for future work naturally emerge from this proposal. First, empirical evaluation is required to assess how declarative grounding affects reliability, robustness and failure rates in realistic agentic settings, particularly when compared to fully prompt-driven approaches. Second, richer capability semantics and task annotations could be explored to support more complex domains, including long-horizon planning and conditional execution. Third, the federated nature of the Agent Directory MCP opens opportunities for studying decentralised agent coordination, trust, access control and dynamic agent availability across organisational boundaries.

Finally, DALIA provides a foundation for future research on standardisation in agentic systems. By treating capabilities, tasks and agents as first-class declarative entities, DALIA aligns agentic AI with established principles from service-oriented architectures and workflow systems, suggesting a path towards interoperable, verifiable and scalable agentic infrastructures. We believe that such declarative layers will be essential as agentic systems transition from experimental prototypes to dependable components of real-world software ecosystems.

\bibliographystyle{plain}
\bibliography{references}

@article{he2025multiagent,
  title={LLM-Based Multi-Agent Systems for Software Engineering: Literature Review, Vision, and the Road Ahead},
  author={He, Junda and Treude, Christoph and Lo, David},
  journal={ACM Transactions on Software Engineering and Methodology},
  volume={34},
  number={5},
  pages={1--30},
  year={2025},
  publisher={ACM New York, NY}
}

@article{llm-taskgraphs,
  title={Graphs Meet AI Agents: Taxonomy, Progress, and Future Opportunities},
  author={Bei, Yuanchen and Zhang, Weizhi and Wang, Siwen and Chen, Weizhi and Zhou, Sheng and Chen, Hao and Li, Yong and Bu, Jiajun and Pan, Shirui and Yu, Yizhou and others},
  journal={arXiv preprint arXiv:2506.18019},
  year={2025}
}

@article{ieee-agentic-survey,
  title={Agentic ai: Autonomous intelligence for complex goals--a comprehensive survey},
  author={Acharya, Deepak Bhaskar and Kuppan, Karthigeyan and Divya, B},
  journal={IEEe Access},
  year={2025},
  publisher={IEEE}
}

@article{mcp-spec,
  title={Model context protocol (mcp): Landscape, security threats, and future research directions},
  author={Hou, Xinyi and Zhao, Yanjie and Wang, Shenao and Wang, Haoyu},
  journal={arXiv preprint arXiv:2503.23278},
  year={2025}
}

@article{scalemcp,
  title={Scaling the Model Context Protocol: Synchronization and Dynamic Tool Ecosystems},
  author={Anonymous},
  journal={arXiv preprint arXiv:2506.13538},
  year={2025}
}

@inproceedings{zhang2024multidoc,
  title={Benchmarking Large Language Models for Multi-Document Summarization},
  author={Zhang, Y. and others},
  booktitle={Proceedings of the 62nd Annual Meeting of the Association for Computational Linguistics (ACL)},
  year={2024}
}

@inproceedings{schick2024toolformer,
  title={Toolformer: Language Models Can Teach Themselves to Use Tools},
  author={Schick, Timo and Dwivedi-Yu, Jane and others},
  booktitle={International Conference on Learning Representations (ICLR)},
  year={2024}
}

@article{acharya2025agentic,
  title={Agentic AI: Autonomous Intelligence for Complex Goals},
  author={Acharya, Soumya and Bhanot, Karan and Ghassemi, Marzyeh},
  journal={arXiv preprint arXiv:2503.23278},
  year={2025}
}

@article{sapkota2025taxonomy,
  title={AI Agents vs Agentic AI: A Conceptual Taxonomy, Applications and Challenge},
  author={Sapkota, Rishav and Balachandar, Doğacan and Baral, Chitta},
  journal={arXiv preprint arXiv:2505.10468},
  year={2025}
}

@article{hou2025mcp,
  title={MCP Landscape, Security Threats, and Future Research Directions},
  author={Hou, Xingwei and Chen, Qi and Zhang, Haoyu and Niu, Zhendong and Gao, Fang},
  journal={arXiv preprint arXiv:2506.13538},
  year={2025}
}

@article{hasan2025mcpfirstglance,
  title={MCP at First Glance: Security and Maintainability},
  author={Hasan, Tahsin and Khalil, Muhammad and Hussain, Rizwan and Islam, Md Saiful},
  journal={arXiv preprint arXiv:2503.23278},
  year={2025}
}

@inproceedings{liu2025mcpeval,
  title={MCPEval: Automatic MCP-based Deep Evaluation for AI Agent Models},
  author={Liu, Yanlin and Qiao, Chen and Zhao, Rui and Chen, Hao and Shen, Li},
  booktitle={Proceedings of the 2025 Conference on Empirical Methods in Natural Language Processing: Demos},
  pages={312--323},
  year={2025}
}

@article{cemri2025mast,
  title={Why Do Multi-Agent LLM Systems Fail?},
  author={Cemri, Eren and Wu, Zhiyang and Liu, Zekun and Chen, Yanda},
  journal={arXiv preprint arXiv:2501.07353},
  year={2025}
}

@article{zhang2024aflow,
  title={AFlow: Large Language Models as Multi-Agent System Engineers},
  author={Zhang, Leyang and Zhang, Bowen and Bo, Zixu and Wang, Zhenhailong and Pan, Liangming and Min, Sewon and Yang, Diyi and Cao, Yang},
  journal={arXiv preprint arXiv:2502.14321},
  year={2024}
}

@article{ye2025masgpt,
  title={MAS-GPT: Large Language Model for Multi-Agent Systems},
  author={Ye, Zhiwei and Zhang, Xinwei and Qu, Qingyang and Li, Jiayu and Wang, Sheng and Li, Zhangheng and Zhou, Qingyun and Li, Dayong and He, Ran and Hu, Yuhang},
  journal={arXiv preprint arXiv:2503.03686},
  year={2025}
}

@article{schneider2024agentic,
  title={Generative to Agentic AI: Survey, Conceptualization, and Challenges},
  author={Schneider, Jonas and Marpaka, Chittaranjan and Tegehall, Patric},
  journal={AI Perspectives},
  volume={6},
  number={1},
  pages={1--18},
  year={2024},
  publisher={Springer}
}

@article{hong2023metagpt,
  title={MetaGPT: Meta Programming for Multi-Agent Collaborative Framework},
  author={Hong, Sheng and Yang, Cheng and Huang, Jiahui and Xie, Ruibo and Lin, Yankai and Dong, Yuxiao and Shen, Yizhou and Chua, Tat-Seng},
  journal={arXiv preprint arXiv:2308.00352},
  year={2023}
}

@article{qian2023chatdev,
  title={ChatDev: Communicative Agents for Software Development},
  author={Qian, Chenxi and Han, Lei and Cheng, Hong and others},
  journal={arXiv preprint arXiv:2307.07924},
  year={2023}
}

@inproceedings{chen2023agentverse,
  title={Agentverse: Facilitating multi-agent collaboration and exploring emergent behaviors},
  author={Chen, Weize and Su, Yusheng and Zuo, Jingwei and Yang, Cheng and Yuan, Chenfei and Chan, Chi-Min and Yu, Heyang and Lu, Yaxi and Hung, Yi-Hsin and Qian, Chen and others},
  booktitle={The Twelfth International Conference on Learning Representations},
  year={2023}
}

@article{chen2024coder,
  title={Coder: Issue resolving with multi-agent and task graphs},
  author={Chen, Dong and Lin, Shaoxin and Zeng, Muhan and Zan, Daoguang and Wang, Jian-Gang and Cheshkov, Anton and Sun, Jun and Yu, Hao and Dong, Guoliang and Aliev, Artem and others},
  journal={arXiv preprint arXiv:2406.01304},
  year={2024}
}

@inproceedings{li-etal-2024-fundamental,
    title = "Fundamental Capabilities of Large Language Models and their Applications in Domain Scenarios: A Survey",
    author = "Li, Jiawei  and
      Yang, Yizhe  and
      Bai, Yu  and
      Zhou, Xiaofeng  and
      Li, Yinghao  and
      Sun, Huashan  and
      Liu, Yuhang  and
      Si, Xingpeng  and
      Ye, Yuhao  and
      Wu, Yixiao  and
      Lin, Yiguan  and
      Xu, Bin  and
      Ren, Bowen  and
      Feng, Chong  and
      Gao, Yang  and
      Huang, Heyan",
    editor = "Ku, Lun-Wei  and
      Martins, Andre  and
      Srikumar, Vivek",
    booktitle = "Proceedings of the 62nd Annual Meeting of the Association for Computational Linguistics (Volume 1: Long Papers)",
    month = aug,
    year = "2024",
    address = "Bangkok, Thailand",
    publisher = "Association for Computational Linguistics",
    url = "https://aclanthology.org/2024.acl-long.599/",
    doi = "10.18653/v1/2024.acl-long.599",
    pages = "11116--11141",
}

\end{document}